%% file: main.tex
\documentclass[conference,onecolumn,draftclsnofoot,letterpaper]{IEEEtran}

\usepackage[utf8]{inputenc} 
\usepackage[T1]{fontenc}
\usepackage{url}
\usepackage{ifthen}
\usepackage[cmex10]{amsmath} 

\usepackage{amsfonts,amssymb,mathrsfs,mathdots,mathtools}

\usepackage{graphicx}
\usepackage{dirtytalk}
\usepackage{color,soul}
\usepackage{dirtytalk}
\usepackage{dsfont}

\usepackage{cite}

\usepackage{caption}
\usepackage{subcaption}

\newtheorem{theorem}{Theorem}
\newtheorem{corollary}[theorem]{Corollary}
\newtheorem{lemma}[theorem]{Lemma}
\newtheorem{definition}[theorem]{Definition}
\newtheorem{prop}[theorem]{Proposition}
\newtheorem{remark}[theorem]{Remark}
\newtheorem{example}[theorem]{Example}

\DeclarePairedDelimiter\abs{\lvert}{\rvert}%
\newcounter{relctr} 
\everydisplay\expandafter{\the\everydisplay\setcounter{relctr}{0}} 

\newcommand\labelrel[2]{%
  \begingroup
    \refstepcounter{relctr}%
    \stackrel{\textnormal{(\alph{relctr})}}{\mathstrut{#1}}%
    \originallabel{#2}%
  \endgroup
}
\AtBeginDocument{\let\originallabel\label} 

\begin{document}
\title{Optimal Maximal Leakage-Distortion Tradeoff} 



\author{%
  \IEEEauthorblockN{Sara~Saeidian\IEEEauthorrefmark{1},
                    Giulia~Cervia\IEEEauthorrefmark{2},
                    Tobias~J.~Oechtering\IEEEauthorrefmark{1},
                    and Mikael Skoglund\IEEEauthorrefmark{1}}
  \IEEEauthorblockA{\IEEEauthorrefmark{1}%
                    KTH Royal Institute of Technology,
                    100 44 Stockholm, Sweden,
                    \{saeidian, oech, skoglund\}@kth.se}
  \IEEEauthorblockA{\IEEEauthorrefmark{2}%
                    IMT Lille Douai,
                    F-59000 Lille, France,
                    giulia.cervia@imt-lille-douai.fr}
}

\maketitle

\begin{abstract}
Most methods for publishing data with privacy guarantees introduce randomness into datasets which reduces the utility of the published data. In this paper, we study the privacy-utility tradeoff by taking maximal leakage as the privacy measure and the expected Hamming distortion as the utility measure. We study three different but related problems. First, we assume that the data-generating distribution (i.e., the prior) is known, and we find the optimal privacy mechanism that achieves the smallest distortion subject to a constraint on maximal leakage. Then, we assume that the prior belongs to some set of distributions, and we formulate a min-max problem for finding the smallest distortion achievable for the worst-case prior in the set, subject to a maximal leakage constraint. Lastly, we define a partial order on privacy mechanisms based on the largest distortion they generate. Our results show that when the prior distribution is known, the optimal privacy mechanism fully discloses symbols with the largest prior probabilities, and suppresses symbols with the smallest prior probabilities. Furthermore, we show that sets of priors that contain more uniform distributions lead to larger distortion, while privacy mechanisms that distribute the privacy budget more uniformly over the symbols create smaller worst-case distortion.
\end{abstract}


\section{Introduction}
\label{sec:intro}
\input{sections/introduction.tex}

\section{Preliminaries and Notation}
\label{sec:background}
\input{sections/background}

\section{Known Prior Distribution}
\label{sec:prob1}
\input{sections/prob1}

\section{A Set of Prior Distributions}
\label{sec:prob2}
\input{sections/prob2}

\section{Ordering Privacy Mechanisms in Terms of Worst-case Distortion}
\label{sec:prob3}
\input{sections/prob3}

\section{Conclusions}
\label{sec:conclusion}
\input{sections/conclusion}

\bibliographystyle{IEEEtran}
\bibliography{refs.bib}

\end{document}

%% file: sections/introduction.tex
How to publish sensitive data safely? This is a question encountered by many data curators. On the one hand, thanks to the rapid progress in big data technologies, it is now possible to extract valuable information from datasets, leading to numerous applications in areas such as image and speech recognition technologies, fraud detection schemes, spam filters, and more. On the other hand, depending on the nature of the data, it may also be possible to extract sensitive information from datasets which raises privacy concerns in data publishing. For instance, it may be inferrable from a person's financial transactions that they have diabetes if they regularly buy insulin. A prime example of this tradeoff concerns health data, where data analysis methods can provide invaluable insights for detecting/treating disorders or the planning of healthcare resources. However, due to the very sensitive nature of health data, a breach of privacy may have severe consequences for the involved participants. 

The most commonly used methods for publishing sensitive data with privacy guarantees rely on privacy mechanisms that introduce randomness into datasets. While randomizing datasets can to some extent alleviate privacy concerns by providing plausible deniability, it may also partly destroy the useful information in the published data. In the privacy literature, this problem is usually referred to as the privacy-utility tradeoff. Roughly speaking, stricter privacy guarantees require increased randomization of the original data, which in turn leads to less utility in the published data. 

In this paper, we will study the privacy-utility tradeoff, where we use the notion of maximal leakage~\cite{braun2009quantitative, issa2019operational} to measure the amount of information leaking through a privacy mechanism. Maximal leakage is an operationally meaningful privacy measure: It captures the inference capabilities of an adversary who observes the published data and tries to guess some (discrete) function of the original data. Furthermore, maximal leakage satisfies a number properties that make it a suitable choice as a privacy metric. For example, no post-processing of the published data can undermine the initial privacy guarantee (i.e., maximal leakage satisfies a data processing inequality)~\cite{issa2019operational}. In addition, maximal leakage can be employed as a tool for studying the privacy guarantees of practical algorithms~\cite{sara2021}. In order to measure the utility of a privacy mechanism, we will use the expected Hamming distortion incurred by the mechanism. Hamming distortion is a commonly used utility measure for discrete data, which is the setting considered in this paper.

Many previous works have studied the privacy-utility tradeoff using different notions of privacy and different utility measures. To give a few relevant examples, taking (local) differential privacy~\cite{dwork2014algorithmic} as the privacy measure, the privacy-utility tradeoff is investigated using Hamming distortion~\cite{sarwate2014rate, wang2016relation, kalantari2018robust}, Bayes risk~\cite{alvim2011differential}, minimax risk~\cite{duchi2013local}, and a class of convex utility functions~\cite{kairouz2016extremal}. In~\cite{liao2018privacy}, maximal $\alpha$-leakage (a generalization of maximal leakage) is taken as the privacy measure, and the privacy-distortion tradeoff is studied using a hard distortion measure which bounds the distortion with probability one. In~\cite{rassouli2019optimal}, total variation privacy is considered and the privacy-utility tradeoff is investigated using mutual information, error probability and mean square error as the utility measures. 

Perhaps the most similar previous work to our current work is \cite{kalantari2018robust}, where the privacy-distortion tradeoff is investigated under local differential privacy and Hamming distortion. In~\cite{kalantari2018robust}, the authors consider a setup in which the distribution generating the original dataset (i.e., the prior distribution) is not exactly known. Instead, we are given a set of distributions such that any member of this set can be the true prior. The authors then categorize sets of distributions into three classes. Class I sets are those sets whose convex hull includes the uniform distribution. Class II sets contain distributions that have the same order in the probabilities assigned to the elements of a given alphabet. To illustrate this, let $\pi = (\pi_1, \pi_2, \pi_3)$ denote a probability distribution on an alphabet with three elements. Then, the sets $\{\pi : \pi_2 \geq \pi_1 \geq \pi_3 \}$ and $\{\pi : \pi_3 \geq \pi_2 \geq \pi_1 \}$ are both examples of Class II sets. Lastly, Class III sets are those sets that belong to neither of the two previous classes. The authors then study the problem of finding the smallest privacy leakage achievable for the worst distribution in a particular set of priors, subject to a bound on the expected distortion. This problem is considered separately for each class of sets of priors. 

Our work is similar to~\cite{kalantari2018robust} in that we are also interested in a robust characterization of the privacy-distortion tradeoff, albeit using a different notion of privacy. More specifically, we have the following setup: Suppose we want to disclose some data, represented as the outcome of a random variable, and subject to a bound on the information leakage. Among all the privacy mechanisms that satisfy the information leakage constraint, we wish to pick the mechanism that incurs the smallest expected distortion. In this setup, we consider three different but related problems:  
\begin{itemize}
    \item Given a fixed prior distribution, what is the smallest expected distortion achievable, subject to an upper bound on the information leakage? What is the optimal privacy mechanism? 
    \item Given a set of priors, what is the smallest expected distortion for the worst-case prior in the set, subject to an upper bound on the information leakage? What is the optimal privacy mechanism? 
    \item Given two privacy mechanisms satisfying the information leakage constraint, and considering the set of all priors, which mechanism can produce larger distortion? 
\end{itemize}
In our study of the second problem, we consider three sub-problems. First, we assume that the set of priors includes the uniform distribution. Then, we relax this assumption, and assume that the set of priors includes a distribution which we will call a \emph{least-informative} distribution. Informally, the least-informative distribution is the most \say{uniform-like} distribution in the set (we will formally define this later). Lastly, we will consider arbitrary sets of priors.

Hence, our approach differs from~\cite{kalantari2018robust} in a few key aspects. First, we will argue that it is not necessary to consider different classes of sets of priors depending on the order of the probabilities assigned to elements in the alphabet. That is, we need not distinguish between Class II and Class III sets. Roughly speaking, this is because maximal leakage does not depend on the labels of the input and output alphabets, and therefore, we can re-label both alphabets without affecting the privacy guarantee of a mechanism. While in our work we consider maximal leakage as our privacy metric, the same argument applies to analysis using local differential privacy since the guarantees of local differential privacy also remain unaffected by re-labelling of the input/output alphabets. 

Another major difference between our work and previous works is that our objective goes well beyond finding the optimal privacy mechanism and characterizing the smallest expected distortion. Here, our goal is to \say{order} both prior distributions and privacy mechanisms based on the utility they provide. For this, we will use methods from majorization theory~\cite{Marshall2011}, which allow us to partially order vectors. In doing so, we will show that, roughly speaking, priors which are more uniformly distributed incur larger expected distortion, while privacy mechanisms that distribute the privacy budget more uniformly over the symbols create smaller worst-case distortion.

The remainder of this paper is organized as follows. In Section~\ref{sec:background}, we will go through some definitions/results related to maximal leakage, Hamming distortion, and majorization theory. We will also define some notations used in the paper. In Section~\ref{sec:prob1}, we will consider the problem of finding the optimal privacy mechanism in the sense of minimizing the expected distortion for a given prior and subject to a constraint on the maximal leakage of the mechanism. In Section~\ref{sec:prob2}, we will generalize the previous scenario by assuming that the prior is not known, but belongs to some fixed set of distributions. Here, we will consider the problem of finding the optimal privacy mechanism for the worst-case prior in the set. Section~\ref{sec:prob3}, concerns a slightly different problem. We assume that we are given two privacy mechanisms satisfying the maximal leakage constraint, and we compare the largest distortion generated by them. Section~\ref{sec:conclusion} presents our conclusions. 

%% file: sections/background.tex
\subsection{Maximal Leakage}
Suppose $X$ is a random variable taking values in a finite alphabet $\mathcal X$. We will use $X$ to represent some sensitive data that we want to publish. In order to release a sanitized version of $X$, we will use the privacy mechanism $P_{Y \mid X}$, which is a conditional probability kernel. This produces a random variable $Y$ taking values in a finite alphabet $\mathcal Y$, which represents the published data. 

Let $\mathcal L(X \to Y)$ denote the maximal leakage from $X$ to $Y$. It is shown in~\cite[Thm. 1]{issa2019operational} that for finite alphabets maximal leakage takes the form
\begin{equation}
\label{eq:maximal_leakage}
    \mathcal{L}(X \to Y) = \log \sum_{y \in \mathcal{Y}} \max_{x \in \mathcal{X} : P_X(x)>0} P_{Y \mid X} (y \mid x), 
\end{equation}
where $\log$ denotes the natural logarithm. From~\eqref{eq:maximal_leakage}, it is clear that maximal leakage depends on the distribution over $X$ (i.e., the prior) only through its support. Therefore, if we fix the support of $X$, we can view maximal leakage as a property of the privacy mechanism $P_{Y \mid X}$. Hence, for a fixed support of $X$, we will adopt the notation $\mathcal L(P_{Y \mid X}) \coloneqq \mathcal{L}(X \to Y)$, and write
\begin{equation}
    \mathcal L(P_{Y \mid X}) = \log \sum_{j=1}^m \max_{i \in [n]} \; P_{Y \mid X}(y_j \mid x_i), 
\end{equation}
where $\abs{\mathrm{supp}(X)} = \abs{\mathcal X} = n \geq 2$, and $\abs{\mathcal Y} = m$. In the above definition, we can interpret $\exp(\mathcal L(P_{Y \mid X}))$ as the overall privacy cost and $\max_{i \in [n]} \, P_{Y \mid X}(y_j \mid x_i)$ as the privacy cost caused by disclosing the $j$th output symbol. We will frequently refer back to this interpretation. In addition, the following upper and lower bounds on maximal leakage, proved in \cite[Lem. 1]{issa2019operational}, will be repeatedly used in the rest of the paper: 
\begin{equation}
\label{eq:maximal_leakage_bounds}
    0 \leq \mathcal L(P_{Y \mid X}) \leq \min \{\log n, \log m\},  
\end{equation}
where the lower bound holds with equality when $X$ and $Y$ are independent, and the upper bound holds with equality when $Y$ is obtained from $X$ through a deterministic mapping.

Let $\Delta^{(m-1)}$ denote the $m-1$-dimensional probability simplex. We will use $\mathcal M^{n,m}$ to denote the set of all $n \times m$ row-stochastic matrices, that is, matrices whose rows are elements from $\Delta^{(m-1)}$. To simplify the notation, a privacy mechanism $P_{Y \mid X}$ will be represented by a row-stochastic matrix $P = [p_{ij}] \in \mathcal M^{n,m}$, where $p_{ij} = P_{Y \mid X}(y_j \mid x_i)$ for $i \in [n]$ and $j \in [m]$. Using this notation, the maximal leakage of a privacy mechanism $P$ can be written as
\begin{equation}
    \mathcal L(P) = \sum_{j=1}^m \max_{i \in [n]} \, p_{ij}, 
\end{equation}
that is, maximal leakage is calculated as the sum of the largest elements in each column of $P$. 

\subsection{Hamming Distortion}
In order to measure the utility of a mechanism $P_{Y \mid X}$, we will calculate the expected distortion $\mathbb E[d(X, Y)]$ using Hamming distortion defined as $d(x, y) = \mathds{1}(x \neq y)$, where $\mathds{1}(\cdot)$ denotes the indicator function. Let $\pi = (\pi_1, \ldots, \pi_n)$ denote the prior distribution on $X$, where $\pi_i$ denotes the probability of $x_i$. Note that we require $\pi_i > 0$ for all $i \in [n]$ since we are assuming that $X$ has full support. The expected Hamming distortion can be written as 
\begin{align}
\label{eq:expected_hamming_distortion}
\begin{split}
    \mathbb E_{P, \pi}[d(X, Y)] &= \sum_{j = 1}^m \sum_{i=1}^n p_{ij} \pi_i \; \mathds{1}(x_i \neq y_j) \\
    &= 1 - \sum_{j = 1}^m \sum_{i=1}^n p_{ij} \pi_i \; \mathds{1}(x_i = y_j).
\end{split}
\end{align}
From~\eqref{eq:expected_hamming_distortion}, it is easy to see that mechanisms with $m > n$, i.e., matrices with more columns than rows, cannot be optimal in terms of minimizing the expected distortion since for $j > n$, we have $\mathds{1}(x_i = y_j) = 0$ for all $i \in [n]$ (this is formally proved in~\cite[Lem. 3]{kalantari2018robust} using local differential privacy, and similar arguments can be made for our case). Hence, in the rest of the paper, we will assume that $m=n$. Note that this also includes the case $m < n$ by having columns in matrix $P$ that consist only of zeros. Therefore, the expected distortion can be written as 
\begin{equation}
\label{eq:expected_hamming_distortion2}
    \mathbb E_{P, \pi}[d(X, Y)] = 1 - \sum_{j = 1}^n  p_{jj} \, \pi_j.
\end{equation}

\subsection{Vector Notations}
\label{ssec:notation}
Consider some vector $x = (x_1, \ldots, x_n) \in \mathbb R^n$. We will be using the following notations to represent a few simple operations on vectors: 
\begin{itemize}
    \item $x_\downarrow = (x_{[1]} , \ldots, x_{[n]})$ denotes a permutation of $x$ that orders it decreasingly, where $x_{[k]}$ is the $k$th largest element in $x$. We will call $x_\downarrow$ the decreasing rearrangement of $x$. 
    \item $x_\uparrow = (x_{(1)} , \ldots, x_{(n)})$ denotes the increasing rearrangement of $x$, where $x_{(k)}$ denotes the $k$th smallest element in $x$. 
    \item $\Tilde x_k = \sum_{j=1}^k x_j$ denotes the sum of the first $k$ elements in $x$.
    \item $\Tilde x_{[k]} = \sum_{j=1}^k x_{[j]}$ denotes the sum of the $k$ largest elements in $x$.
    \item $\Tilde x_{(k)} = \sum_{j=1}^k x_{(j)}$ denotes the sum of the $k$ smallest elements in $x$.
    \item $i_x = (i_x(1), \ldots, i_x(n))$ denotes the sequence of indexes corresponding to the decreasing rearrangement of $x$, that is, $x_\downarrow = (x_{[1]} , \ldots, x_{[n]}) = (x_{i_x(1)} , \ldots, x_{i_x(n)})$, where $i_x(j) \in [n]$ is the index of the $j$th largest element in $x$. 
\end{itemize}

\subsection{Majorization}
\begin{figure}
     \centering
     \begin{subfigure}[b]{0.48\textwidth}
        \captionsetup{justification=centering}
        \includegraphics[scale=0.2]{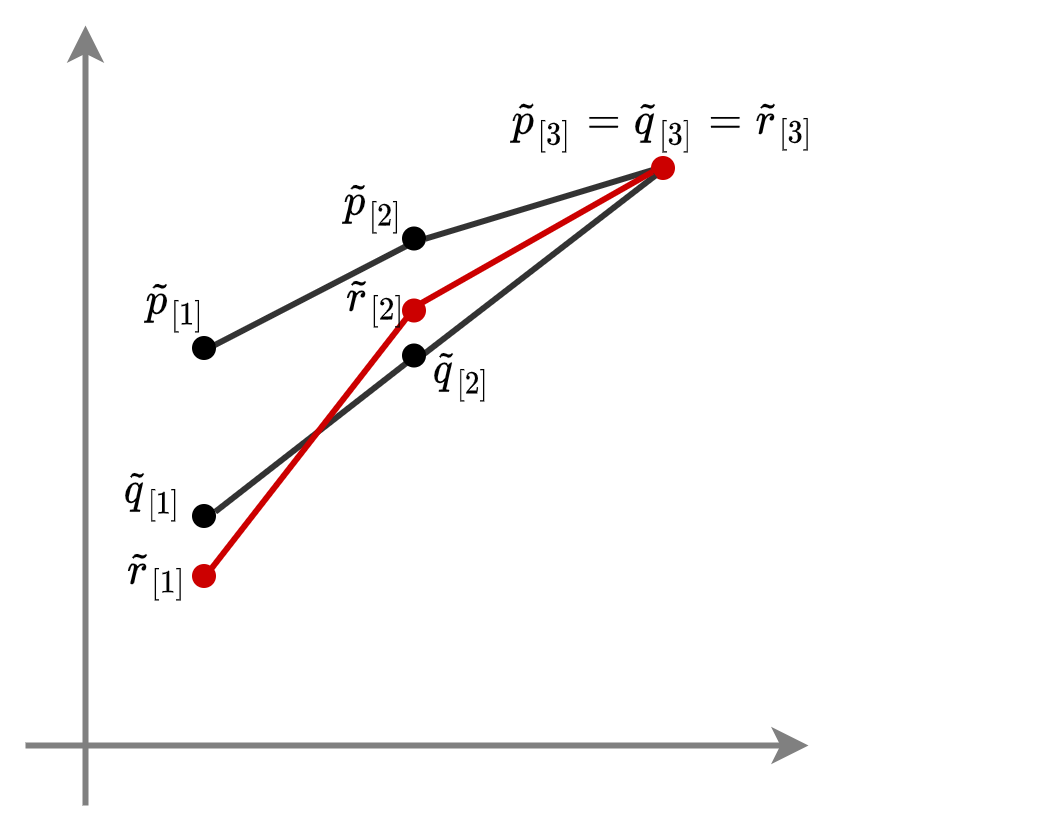}
        \caption{$\tilde{p}_{[k]} > \tilde{q}_{[k]}, \tilde{r}_{[k]}$ for $k=1,2$ and $\tilde{p}_{[3]} = \tilde{q}_{[3]} = \tilde{r}_{[3]}$. Since $\tilde{q}_{[1]} > \tilde{r}_{[1]} $ but $\tilde{q}_{[2]} < \tilde{r}_{[2]}$, $q$ and $r$ cannot be compared.}
     \end{subfigure}
     \hfill
     \begin{subfigure}[b]{0.48\textwidth}
         \captionsetup{justification=centering}
         \includegraphics[scale=0.2]{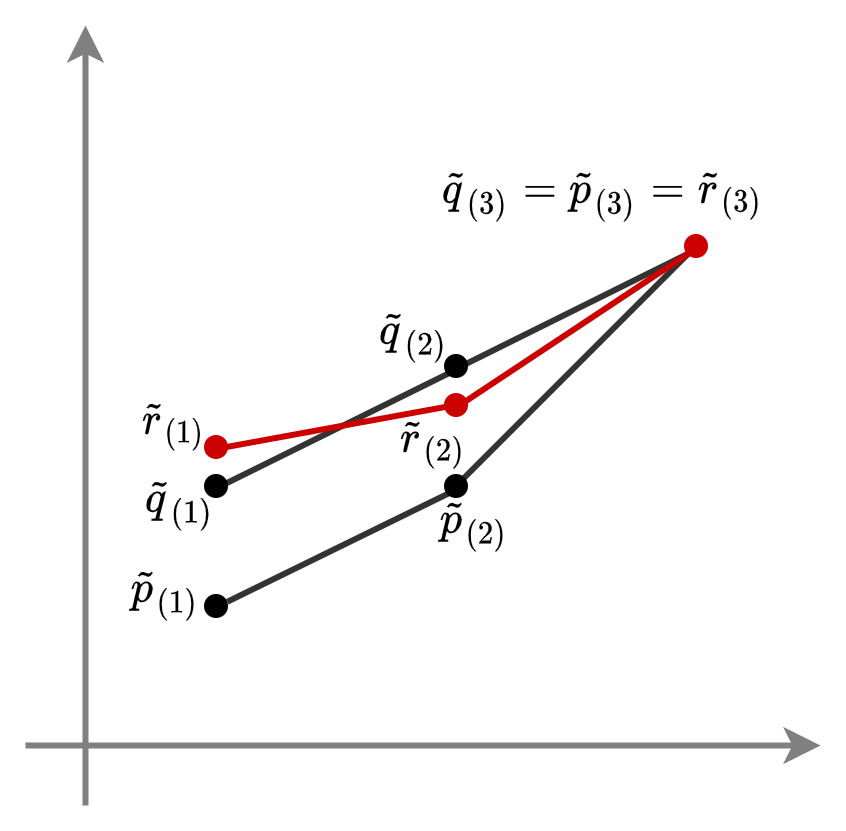}
         \caption{$\tilde{p}_{(k)} \! < \!\tilde{q}_{(k)}, \tilde{r}_{(k)}$ for $k=1,2$ and $\tilde{p}_{(3)} \!= \! \tilde{q}_{(3)} \!=\! \tilde{r}_{(3)}$. Since $\tilde{q}_{(1)} < \tilde{r}_{(1)} $ but $\tilde{q}_{(2)} > \tilde{r}_{(2)}$, $q$ and $r$ cannot be compared.}
     \end{subfigure}
        \caption{Illustration of majorization using three vectors $p,q,r \in \mathbb{R}^3_+$, where we have $q,r \prec p$, but $q$ and $r$ cannot be compared in terms of majorization.}
        \label{fig:majorization}
\end{figure}

In this section, we will give a few definitions and results from majorization theory. Informally, majorization theory allows us to order vectors based on how \say{uniform} the elements of the vectors are. All of the following definitions/results can be found in~\cite{Marshall2011}. 
\begin{definition}[Majorization]\label{def:majorization}
Consider two vectors $p, q \in \mathbb{R}^n$. We say that $p$ majorizes $q$, and write $q \prec p$ if 
\begin{equation}
\Tilde q_{[m]} \leq \Tilde p_{[m]} \quad \text{for} \quad m = 1, \ldots, n-1 \quad \text{and} \quad \Tilde p_{n} = \Tilde q_{n},
\end{equation}
or alternatively, 
\begin{equation}
\Tilde q_{(m)} \geq \Tilde p_{(m)} \quad \text{for} \quad m = 1, \ldots, n-1 \quad \text{and} \quad \Tilde p_{n} = \Tilde q_{n}.
\end{equation}
\end{definition}
Majorization defines a partial order on $n$-dimensional vectors, i.e., a relation that is reflexive, transitive, and anti-symmetric. Note that not all $n$-dimensional vectors can be compared in terms of majorization. For example, $(4,4,1)$ and $(5,2,2)$ cannot be compared in terms of majorization. On the other hand, if we define $\mathcal{Q} = \{(q_1, q_2, q_3) \in \mathbb{R}_+^3 : \sum_{i=1}^3 q_i = 9\}$, then $(3,3,3)$ is majorized by all $q \in \mathcal{Q}$ while $(9,0,0)$, $(0,9,0)$ and $(0,0,9)$ majorize all $q \in \mathcal{Q}$. A graphical illustration of majorization is given in Figure~\ref{fig:majorization}. 

\begin{figure}
     \centering
     \begin{subfigure}[t]{0.48\textwidth}
        \includegraphics[scale=0.2]{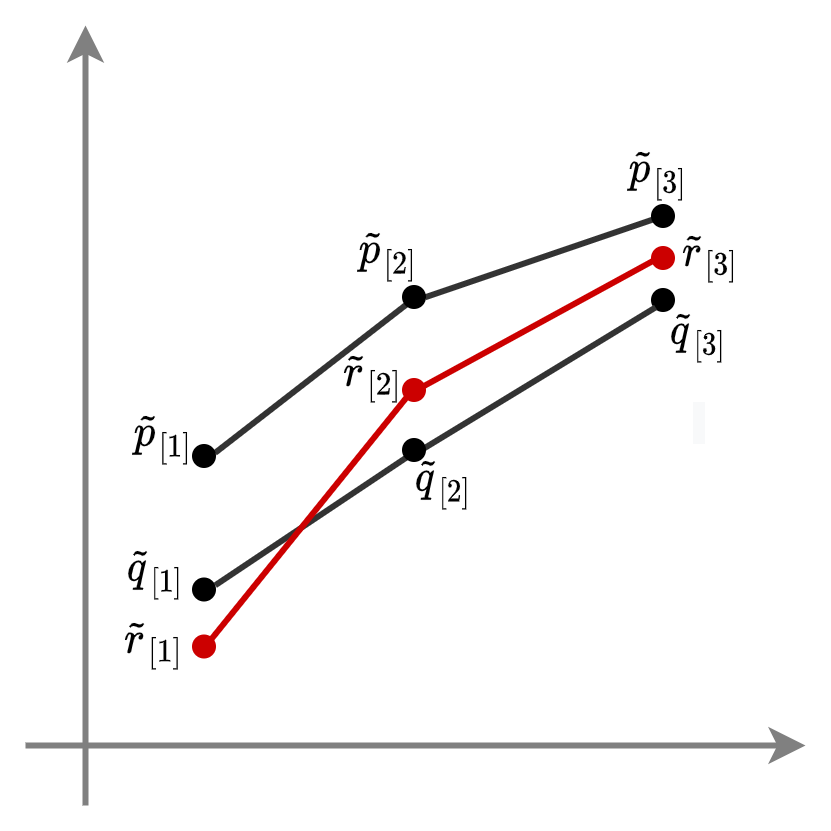}
        \caption{$q,r \prec_w p$ but $q$ and $r$ cannot be compared.}
     \end{subfigure}
     \hfill
     \begin{subfigure}[t]{0.48\textwidth}
         \centering
         \includegraphics[scale=0.2]{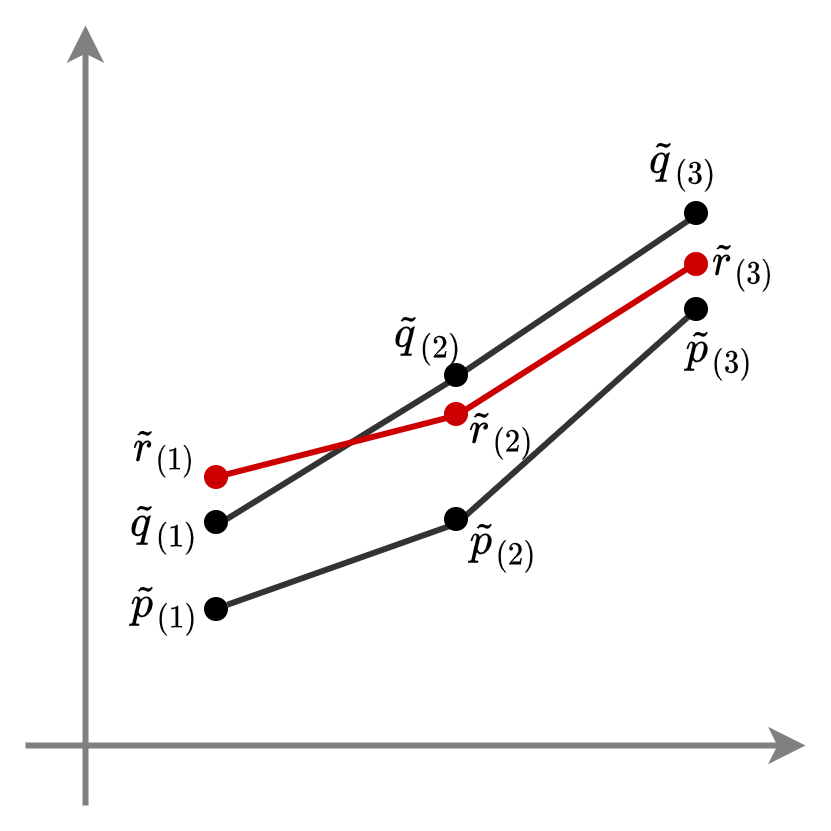}
         \caption{$q,r \prec^w p$ but $q$ and $r$ cannot be compared.}
     \end{subfigure}
        \caption{Illustration of weak majorization using three vectors $p,q,r \in \mathbb{R}^3_+$.}
        \label{fig:weak_majorization}
\end{figure}

In Definition~\ref{def:majorization}, the sum of the elements in vectors $p$ and $q$ are required to be equal. If we remove this condition, we will get the following two extensions of majorization, which also define partial orders on vectors.
\begin{definition}[Weak majorization]
Consider two vectors $p, q \in \mathbb{R}^n$. We say that $p$ weakly sub-majorizes $q$, and write $q \prec_w p$ if 
\begin{equation}
\Tilde q_{[m]} \leq \Tilde p_{[m]} \quad \text{for all} \quad m = 1, \ldots, n.
\end{equation}
Furthermore, we say that $p$ weakly super-majorizes $q$, and write $q \prec^w p$ if 
\begin{equation}
\Tilde q_{(m)} \geq \Tilde p_{(m)} \quad \text{for all} \quad m = 1, \ldots, n.
\end{equation}
\end{definition}
Weak majorization is illustrated in Figure~\ref{fig:weak_majorization}. 

\begin{definition}[Schur-convex function]
Consider two vectors $p, q \in \mathbb{R}^n$. We say that a real-valued function $\Phi: \mathbb{R}^n \to \mathbb R$ is Schur-convex if $q \prec p$ implies $\Phi(q) \leq \Phi(p)$.
\end{definition}
The previous definition states that Schur-convex functions are order-preserving (i.e., increasing) with respect to the majorization partial order. In the next lemma, we describe a common method for proving Schur-convexity of functions. 

\begin{lemma}[Schur's condition]
\label{lemma:Schur_condition}
Let $I \subset \mathbb R$ be an open interval and suppose $\Phi : I^n \to \mathbb R$ is a continuously differentiable function. Then, necessary and sufficient conditions for $\Phi$ to be Schur-convex are 
\begin{equation}
    \text{$\Phi$ is symmetric on $I^n$}
\end{equation}
and 
\begin{equation}
    (p_i - p_j) \big(\frac{\partial \Phi}{\partial p_i} - \frac{\partial \Phi}{\partial p_j} \big) \geq 0 \quad \text{for all} \quad i,j \in [n].
\end{equation}

\end{lemma}
In order for a Schur-convex to be order-preserving with respect to weak majorization, we need to specify an extra condition on the function. This is illustrated in the following lemma. 
\begin{lemma}
Suppose $\Phi: \mathbb{R}^n \to \mathbb R$ is a Schur-convex function. If $\Phi$ is increasing, then $q \prec_w p$ implies $\Phi(q) \leq \Phi(p)$. Conversely, if $\Phi$ is decreasing, then $q \prec^w p$ implies $\Phi(q) \leq \Phi(p)$. 
\end{lemma}

%% file: sections/prob1.tex
In this section, we study our first problem formulated as follows. Suppose we want to disclose the outcome of a random variable $X$ such that the information leakage about $X$ is below a predefined threshold, and assuming that the prior distribution $\pi$ over $X$ is known. Among all the privacy mechanisms that satisfy the leakage constraint, we want to pick the mechanism that creates the smallest expected Hamming distortion, and therefore, provides the highest utility. Considering this setup, let
\begin{equation}
    \mathcal S_\gamma = \{P \in \mathcal M^{n, n} : \mathcal L(P) \leq \gamma \}
\end{equation}
be the set of all $n \times n$ row-stochastic matrices whose maximal leakage is bounded by some $\gamma \leq \log n$, where $e^\gamma$ represents our overall privacy budget. Note that when $\gamma \geq \log n$, we are allowed to fully disclose the outcomes of $X$, in which case $\mathcal S_\gamma = \mathcal M^{n, n}$, i.e., the set $\mathcal S_\gamma$ contains all $n \times n$ row-stochastic matrices. Thus, in the following we consider $\gamma \leq \log n$. Our goal is to find $D_{\mathrm{min}}(\gamma, \pi)$ defined as 
\begin{equation}
    D_{\mathrm{min}}(\gamma, \pi) \coloneqq \inf_{P \in \mathcal S_\gamma} \mathbb E_{P, \pi}[d(X, Y)] = \inf_{P \in \mathcal S_\gamma} (1 - \sum_j p_{jj} \pi_j) = 1 - \sup_{P \in \mathcal S_\gamma} \sum_j p_{jj} \pi_j. 
\label{eq:problem_fixed_prior}    
\end{equation}

Problem~\eqref{eq:problem_fixed_prior} describes a constrained convex optimization problem: The objective function is linear, and one can easily verify that the set $S_\gamma$ is convex. The following result shows that the optimal mechanism for this problem fully discloses symbols with the largest prior probabilities, and suppresses symbols with the smallest prior probabilities.
\begin{theorem}
\label{theo:opt_channel_fixed_prior}
Suppose $k$ is a positive integer such that $k \leq e^\gamma \leq k+1$ and $k \leq n-1$. Then, the smallest expected distortion in problem~\eqref{eq:problem_fixed_prior} is 
\begin{equation}
    D_\mathrm{min} (\gamma, \pi) = 1 - \Big(\Tilde{\pi}_{[k]} + (e^\gamma - k) \pi_{[k+1]}  \Big). 
\end{equation}
In addition, the optimal privacy mechanism $P^*$ satisfies
\begin{equation}
\label{eq:diagonals_largest}
   \max_{i \in [n]} p_{ij}^* = p_{jj}^*, 
\end{equation}
for all $j \in [n]$ (i.e., the largest element in each column is located on the diagonal), and has the following diagonal entries:  
\begin{equation}
\label{eq:optimal_diagonals}
    p_{jj}^* = \begin{cases}
    1 & j = i_\pi(1), \ldots, i_\pi(k),\\
    e^\gamma - k & j=i_\pi(k+1),\\
    0  & j= i_\pi(k+2), \ldots, i_\pi(n). 
    \end{cases}
\end{equation}
\end{theorem}
\begin{IEEEproof}
Take two vectors $x, y \in \mathbb R^n_+$. We will define a partial order on $\mathbb R^n_+$ induced by $i_\pi$ as follows: 
\begin{equation}
    x \prec_{i_\pi} y \quad \text{if and only if} \quad \sum_{j =1}^l x_{i_\pi(j)} \leq  \sum_{j =1}^l y_{i_\pi(j)},
\end{equation}
for all $l=1, \ldots, n$, where $i_\pi(j)$ is the index of the $j$th largest element in $\pi$ (see Section~\ref{ssec:notation}). Note that this partial order is very similar to the weak sub-majorization order, except the elements in the vectors $x$ and $y$ are ordered according to $i_\pi$ instead of decreasingly. Now, let $p_\mathrm{diag} = (p_{11}, \ldots, p_{nn})$ denote the vector of diagonal entries for matrix $P$. We will use our partial order induced by $i_\pi$ on vectors with non-negative elements to define a pre-order on the matrices in $\mathcal S_\gamma$: for $P, Q \in \mathcal S_\gamma$, we have
\begin{equation}
\label{eq:pre-order}
    P \prec_{i_\pi} Q \quad \text{if and only if} \quad p_{\mathrm{diag}} \prec_{i_\pi} q_{\mathrm{diag}}. 
\end{equation}
Note that $i_\pi$ only induces a pre-order on matrices since the relation in~\eqref{eq:pre-order} is reflexive and transitive but not anti-symmetric. The result stated in the theorem is then immediate by noting that: \\
(a) The function $f_\pi(P) = \sum_j p_{jj} \pi_j$ is order-preserving (i.e., increasing) with respect to the pre-order $\prec_{i_\pi}$, that is,
\begin{equation}
\label{eq:monotone_function_preorder}
    P \prec_{i_\pi} Q \implies f_\pi (P) \leq f_\pi(Q). 
\end{equation}
(b) $\sum_{j=1}^n p_{jj} \leq \sum_{j=1}^n \max_i p_{ij} \leq e^\gamma$, for all $P \in \mathcal S_\gamma$, with equality when $\max_i p_{ij} = p_{jj}$ for all $j \in [n]$ and $\sum_j p_{jj} = e^\gamma$. \\
(c) A matrix $P^*$ described by~\eqref{eq:diagonals_largest} and~\eqref{eq:optimal_diagonals} satisfies $P \prec_{i_\pi} P^*$ for all $P \in \mathcal S_\gamma$.  
\end{IEEEproof}

\begin{remark}
Conditions~\eqref{eq:diagonals_largest} and~\eqref{eq:optimal_diagonals} together imply that for $\gamma$ such that $k < e^\gamma \leq k+1$, the optimal privacy mechanism for problem~\eqref{eq:problem_fixed_prior} has $n - (k+1)$ all-zero columns. Hence, the output alphabet has support of size $k+1$. 
\end{remark}

\begin{remark}
\label{rem:optimal_mech_prior_relation}
The optimal mechanism for problem~\eqref{eq:problem_fixed_prior} depends on the prior only through $i_\pi$. We will frequently use this property in the rest of the paper. 
\end{remark}

In Theorem~\ref{theo:opt_channel_fixed_prior}, if we view $\max_{i \in [n]} p_{ij}$ as the privacy cost of disclosing the $j$th symbol, then the optimal mechanism is highly opportunistic in that the privacy budget is allocated only to the most likely symbols. Note that we must be careful in interpreting this result. While for a fixed prior an opportunistic mechanism is optimal, we cannot conclude that, in general, privacy mechanisms that allocate the privacy budget uniformly to all symbols will generate larger distortion. In fact, we will see in Section~\ref{sec:prob3} that when considering the set of all priors, mechanisms that distribute the privacy budget more uniformly among the symbols generate smaller worst-case distortion. 

%% file: sections/prob2.tex
Now, suppose the prior distribution is not known, but belongs to some set $\Pi$ of probability distributions with support of size $n$ (the largest set $\Pi$ is the relative interior of $\Delta^{(n-1)}$). Our goal is to find a privacy mechanism in $\mathcal S_\gamma$ that minimizes the expected distortion for the worst-case prior in $\Pi$. Thus, the problem is changed to finding $D_\mathrm{min} (\gamma, \Pi)$ defined as 
\begin{align}
\begin{split}
    D_\mathrm{min} (\gamma, \Pi) &\coloneqq \inf_{P \in \mathcal S_\gamma} \; \sup_{\pi \in \Pi} \, \mathbb E_{P, \pi}[d(X, Y)] \\
    &= \inf_{P \in \mathcal S_\gamma} \; \sup_{\pi \in \Pi} (1 - \sum_j p_{jj} \pi_j) \\
    &= 1 - \sup_{P \in \mathcal S_\gamma} \; \inf_{\pi \in \Pi} \sum_j p_{jj} \pi_j. 
\end{split}
\label{eq:problem_set_of_priors}    
\end{align}

We will study this problem by considering three sub-problems: first, we will assume that the set $\Pi$ contains the uniform distribution. Then, we will relax this condition, and assume that $\Pi$ contains a \emph{least-informative} distribution. Informally, one can think of the least-informative distribution as the distribution in $\Pi$ that is more uniform than any other distribution in $\Pi$. Lastly, we will consider an arbitrary set $\Pi$. 
\subsection{Sets Containing the Uniform Distribution}
Suppose the set $\Pi$ contains the uniform distribution. Note that we are not making any other assumptions about $\Pi$ such as convexity, compactness, etc. In this case, we have the following result characterizing $D_\mathrm{min} (\gamma, \Pi)$. 
\begin{prop}
\label{prop:prior_set_contains_uniform}
Suppose the set $\Pi$ contains the uniform distribution denoted by $\pi^u$. Then, the smallest expected distortion for problem~\eqref{eq:problem_set_of_priors} is 
\begin{equation}
\label{eq:optimal_distortion_uniform}
    D_\mathrm{min} (\gamma, \Pi) = 1 - \frac{e^\gamma}{n},
\end{equation}
which is achieved by any privacy mechanism $P \in \mathcal{S}_\gamma$ satisfying $\sum_j p_{jj} = e^\gamma$ with $\pi^u$ as the prior. 
\end{prop}
\begin{IEEEproof}
We prove this result by showing that the RHS of~\eqref{eq:optimal_distortion_uniform} both lower bounds and upper bounds the LHS. 

\underline{Lower bound}: Let $\pi^u$ denote the uniform distribution, that is, $\pi_1^u = \ldots = \pi_n^u = \frac{1}{n}$. Then, for all $P \in \mathcal S_\gamma$ we have 
\begin{align}
    \begin{split}
    \inf_{\pi \in \Pi} \sum_j p_{jj} \pi_j &\leq \sum_j p_{jj} \pi_j^u\\
    &= \frac{1}{n} \sum_j p_{jj}\\
    &\leq \frac{1}{n} \sum_j \max_i p_{ij}\\
    &\leq \frac{e^\gamma}{n}. \\
    \end{split}
\end{align}
Hence, by taking the supremum of both sides we get
\begin{equation}
    \sup_{P \in \mathcal S_\gamma} \; \inf_{\pi \in \Pi} \sum_j p_{jj} \pi_j \leq \frac{e^\gamma}{n}, 
\end{equation}
and finally, 
\begin{equation}
    1 - \sup_{P \in \mathcal S_\gamma} \; \inf_{\pi \in \Pi} \sum_j p_{jj} \pi_j \geq 1 - \frac{e^\gamma}{n}. 
\end{equation}
\underline{Upper bound:} Fix some $Q \in \mathcal S_\gamma$ such that 
\begin{equation}
    \max_i q_{ij} = q_{jj} = \frac{e^\gamma}{n}
\end{equation}
for all $j \in [n]$. Then, we can write 
\begin{equation}
    1 - \sup_{P \in \mathcal S_\gamma} \; \inf_{\pi \in \Pi} \sum_j p_{jj} \pi_j \leq 1 - \inf_{\pi \in \Pi} \sum_j q_{jj} \pi_j = 1 - \frac{e^\gamma}{n}. 
\end{equation}
Finally, we verify that any matrix $P \in \mathcal S_\gamma$ satisfying $\sum_j p_{jj} = e^\gamma$ achieves $D_\mathrm{min} (\gamma, \Pi)$ with $\pi^u$ as the prior: 
\begin{equation}
    1 - \sum_{j} p_{jj} \pi_j^u = 1 - \frac{1}{n} \sum_{j} p_{jj} = 1 - \frac{e^\gamma}{n}. 
\end{equation}
\end{IEEEproof}
Proposition~\ref{prop:prior_set_contains_uniform} suggests that the worst prior in $\Pi$ is in fact the uniform distribution. Therefore, in the next section we will look into sets that do not necessarily contain the uniform distribution, but contain a distribution that is more uniform that any other distribution in the set.  

\subsection{Sets Containing a Least-informative Distribution}
Here, we will relax the condition on $\Pi$ in that we no longer require $\Pi$ to contain the uniform distribution; we only require that $\Pi$ contains a least-informative distribution. For the rest of this section, we will assume $k \leq e^\gamma \leq k+1$ for some positive integer $k \leq n-1$. 

\begin{lemma}
\label{lemma:schur-convexity-prior}
Consider the function $h_\gamma : \mathbb R_+^n \to \mathbb R_+$ defined as $h_\gamma(\pi) \coloneqq \sup_{P \in \mathcal S_\gamma} \sum_j p_{jj} \pi_j$. Then, $h_\gamma$ depends on $\pi$ through the function $f_\gamma : \mathbb R_+^n \to \mathbb R_+^2$ defined as $f_\gamma (\pi) = (\Tilde \pi_{[k]}, \pi_{[k+1]})$. Moreover, $h_\gamma$ is increasing and Schur-convex in $f_\gamma(\pi)$, $\pi \in \Pi$. Thus, for all $\pi, \rho \in \Pi$ such that $f_\gamma(\pi) \prec_w f_\gamma(\rho)$, we have $h_\gamma(\pi) \leq h_\gamma(\rho)$.
\end{lemma}
\begin{IEEEproof}
First, we apply the result of Theorem~\ref{theo:opt_channel_fixed_prior} to get
\begin{equation}
    h_\gamma(\pi) = \sup_{P \in \mathcal S_\gamma} \sum_j p_{jj} \pi_j = \Tilde \pi_{[k]} + (e^\gamma - k) \pi_{[k+1]},
\end{equation}
from which we can see that $h_\gamma$ depends on $\pi$ only through $f_\gamma (\pi) = (\Tilde \pi_{[k]}, \pi_{[k+1]})$.  Now, can prove the Schur-convexity of $h_\gamma$ by verifying Schur's condition (Lemma~\ref{lemma:Schur_condition}). Observe that $h_\gamma$ is symmetric with respect to permutations of $\pi$. This is because maximal leakage $\mathcal L(P)$ does not depend on the order of rows and column of $P$. Therefore, if $P \in \mathcal S_\gamma$, then $TP \in \mathcal{S}_\gamma$, where $T$ is some $n \times n$ permutation matrix. In addition, since
\begin{equation}
    \frac{\partial h_\gamma(\pi)}{\Tilde \pi_{[k]}} \geq \frac{\partial h_\gamma(\pi)}{\pi_{[k+1]}} \geq 0,
\end{equation}
Schur's condition is satisfied and $h_\gamma(\pi)$ is increasing in $f_\gamma(\pi)$. 
\end{IEEEproof}

Lemma~\ref{lemma:schur-convexity-prior} suggests that for calculating $D_\mathrm{min} (\gamma, \Pi)$ one needs to consider the the most uniform prior in $\Pi$. To formalize uniformity, we make the following definition. 

\begin{definition}
Let $f_\gamma$ be the function defined in Lemma~\ref{lemma:schur-convexity-prior}. We say that a distribution $\pi^* \in \Pi$ is $k$-least-informative if it satisfies $f_\gamma(\pi^*) \prec_w f_\gamma(\pi)$ for all $\pi \in \Pi$. 
\end{definition}

\begin{prop}
\label{prop:prior_set_contains_least_informative}
Assume that the set $\Pi$ contains a $k$-least-informative distribution denoted by $\pi^*$. Then, the smallest expected distortion for problem~\eqref{eq:problem_set_of_priors} is 
\begin{equation}
    D_\mathrm{min} (\gamma, \Pi) = 1 - \Big( \Tilde \pi^*_{[k]} + (e^\gamma - k) \pi^*_{[k+1]} \Big).
\end{equation}
Furthermore, $D_\mathrm{min} (\gamma, \Pi)$ is achieved by any stochastic matrix satisfying~\eqref{eq:diagonals_largest} and~\eqref{eq:optimal_diagonals} for prior $\pi^*$. 
\end{prop}
\begin{IEEEproof}
From (the proof of) Lemma~\ref{lemma:schur-convexity-prior}, we know that $\sup_{P \in \mathcal S_\gamma} \sum_{j} p_{jj} \pi_j$ is symmetric in $\pi$. Therefore, it suffices to consider the decreasing rearrangement of the distributions in $\Pi$. That is, we assume $\pi_1 \geq \ldots \geq \pi_n$ for all $\pi \in \Pi$. Now, we will prove an upper bound and a lower bound on $D_\mathrm{min} (\gamma, \Pi)$. 

\underline{Lower bound}: 
\begin{align}
\begin{split}
    \sup_{P \in \mathcal S_\gamma} \; \inf_{\pi \in \Pi} \sum_j p_{jj} \pi_j & \leq \inf_{\pi \in \Pi} \; \sup_{P \in \mathcal S_\gamma} \sum_j p_{jj} \pi_j \\
    &\labelrel={equa} \sum_{j=1}^k \pi^*_j + (e^\gamma - k) \pi^*_{k+1}\\
    &= \Tilde \pi^*_{k} + (e^\gamma - k) \pi^*_{k+1},
\end{split}
\end{align}
where equality~\eqref{equa} follows from Lemma~\ref{lemma:schur-convexity-prior}. Therefore, 
\begin{equation}
    1 - \sup_{P \in \mathcal S_\gamma} \; \inf_{\pi \in \Pi} \sum_j p_{jj} \pi_j \geq 1 - \Big( \sum_{j=1}^k \pi^*_j + (e^\gamma - k) \pi^*_{k+1} \Big). 
\end{equation}
\underline{Upper bound}:  Fix some $P^* \in \mathcal S_\gamma$ satisfying~\eqref{eq:diagonals_largest},~\eqref{eq:optimal_diagonals} for $i_{\pi} = (1, \ldots, n)$. Then, we can write 
\begin{align}
\begin{split}
    1 - \sup_{P \in \mathcal S_\gamma} \; \inf_{\pi \in \Pi} \sum_j p_{jj} \pi_j &\leq 1 - \inf_{\pi \in \Pi} \sum_j p^*_{jj} \pi_j \\
    &= 1 - \inf_{\pi \in \Pi} \Tilde \pi_k + (e^\gamma - k) \pi_{k+1}\\
    &= 1 - \Big( \Tilde \pi^*_{k} + (e^\gamma - k) \pi^*_{k+1}\Big).
\end{split}
\end{align}
\end{IEEEproof}

\begin{remark}
The uniform distribution over an alphabet of size $n$ is $k$-least informative for all $k \leq n-1$. Therefore, Proposition~\ref{prop:prior_set_contains_uniform} can be considered as a special case of Proposition~\ref{prop:prior_set_contains_least_informative}.
\end{remark}

\subsection{General Sets}
In the previous section, we considered sets of priors which contain a least-informative distribution. One should note that, in general, a set $\Pi$ may not contain a least-informative distribution since $\prec_w$ is a partial order, and not all members of $\Pi$ may be comparable in terms of $\prec_w$. To address this matter, in this section we present a general approach for finding $D_\mathrm{min} (\gamma, \Pi)$. 

Let $\Pi_\downarrow$ be the set containing the decreasing rearrangement of the priors in $\Pi$, i.e., $\Pi_\downarrow = \{\pi_\downarrow : \pi \in \Pi \}$. 
\begin{theorem}
\label{theo:general_solution}
For all $\Pi \neq \emptyset$, the smallest expected distortion $D_\mathrm{min} (\gamma, \Pi)$ can be obtained as the solution to the following optimization problem: 
\begin{equation}
\label{eq:general_solution}
    D_\mathrm{min} (\gamma, \Pi) = 1 - \inf_{\pi \in \Pi_\downarrow} \; \Tilde \pi_k + (e^\gamma - k) \pi_{k+1}.
\end{equation}
Furthermore, the optimal privacy mechanism $P \in \mathcal S_\gamma$ for problem~\eqref{eq:general_solution} satisfies~\eqref{eq:diagonals_largest} and~\eqref{eq:optimal_diagonals} for $i_\pi = (1, \ldots, n)$. 
\end{theorem}
\begin{IEEEproof}
First, we argue that in order to solve problem~\eqref{eq:problem_set_of_priors}, we can consider the set $\Pi_\downarrow$ instead of $\Pi$. As stated in (the proof of) Lemma~\ref{lemma:schur-convexity-prior}, the set $\mathcal S_\gamma$ is permutation-invariant.  That is, if $P \in \mathcal S_\gamma$, then $TP \in \mathcal{S}_\gamma$, where $T$ is some $n \times n$ permutation matrix. From this, we conclude that without loss of generality we can order the elements of each $\pi \in \Pi$ in some predefined way, for example, decreasingly. 

Now, we show that the RHS of~\eqref{eq:general_solution} both lower bounds and upper bounds the LHS. 

\underline{Lower bound}:
\begin{align}
\begin{split}
    1 - \sup_{P \in \mathcal S_\gamma} \; \inf_{\pi \in \Pi} \sum_j p_{jj} \pi_j &= 1 - \sup_{P \in \mathcal S_\gamma} \; \inf_{\pi \in \Pi_\downarrow} \sum_j p_{jj} \pi_j\\
    &\geq  1 - \inf_{\pi \in \Pi_\downarrow} \; \sup_{P \in \mathcal S_\gamma} \sum_j p_{jj} \pi_j\\ 
    &= 1 - \inf_{\pi \in \Pi_\downarrow} \; \sum_{j=1}^k \pi_j + (e^\gamma - k) \pi_{k+1},
\end{split}
\end{align}
where the last equality follows from Theorem~\ref{theo:opt_channel_fixed_prior} since $i_\pi = (1, \ldots, n)$ for all $\pi \in \Pi_\downarrow$. 

\underline{Upper bound}: Let $P^* \in \mathcal{S}_\gamma$ be some stochastic matrix satisfying~\eqref{eq:diagonals_largest} and~\eqref{eq:optimal_diagonals} for $i_\pi = (1, \ldots, n)$. Then, we have 
\begin{align}
\begin{split}
    1 - \sup_{P \in \mathcal S_\gamma} \; \inf_{\pi \in \Pi_\downarrow} \sum_j p_{jj} \pi_j &\leq  1 - \inf_{\pi \in \Pi_\downarrow} \sum_j p^*_{jj} \pi_j\\ 
    &= 1 - \inf_{\pi \in \Pi_\downarrow} \; \sum_{j=1}^k \pi_j + (e^\gamma - k) \pi_{k+1}. 
\end{split}
\end{align}
Finally, the fact that the optimal mechanism for problem~\eqref{eq:general_solution} satisfies~\eqref{eq:diagonals_largest} and~\eqref{eq:optimal_diagonals} for $i_\pi = (1, \ldots, n)$ follows from Remark~\ref{rem:optimal_mech_prior_relation}. 
\end{IEEEproof}

Theorem~\ref{theo:general_solution} states that for an arbitrary set $\Pi$, the smallest expected distortion can be calculated as the solution to an optimization problem over two variables. In fact, we can combine the first $k$ elements of $\pi$ into $\Tilde{\pi}_k$ for all $\pi \in \Pi_\downarrow$, and end up with an optimization problem with two variables over a two-dimensional set. This is of course a very convenient property: Regardless of how large $n$ is, we only need to optimize over two variables. 

In the following two numerical examples, we will illustrate the results of this section. 
\begin{example}
Suppose we want to solve problem~\eqref{eq:problem_set_of_priors} with $\gamma = \log 2.5$ and for a set $\Pi^{(1)}$ of distributions over an alphabet with four elements defined as $\Pi^{(1)}=\{\pi \in \Delta^{(3)} : \pi = (0.4 - 2\delta,\, 0.3+\delta,\, 0.15+0.5 \delta,\, 0.15+0.5 \delta),\, 0 \leq \delta \leq 0.1 \}$. To solve the problem, first we need to construct the set $\Pi_\downarrow^{(1)}$. For this, we note that for $0 \leq \delta \leq \frac{1}{30}$ we have 
\begin{equation}
    0.4 - 2\delta \geq  0.3+\delta > 0.15+0.5 \delta,
\end{equation}
while for $\frac{1}{30} \leq \delta \leq 0.1$ we have 
\begin{equation}
    0.3+\delta \geq 0.4 - 2\delta  \geq 0.15+0.5 \delta.
\end{equation}
However, since $k=2$, we can sum over the two largest elements of all $\pi$ to obtain the two dimensional set $\Pi_\downarrow^{(1)} = \{\pi \in \Delta^{(2)} : \pi=(0.7 - \delta,\, 0.15+0.5 \delta,\, 0.15+0.5 \delta),\, 0 \leq \delta \leq 0.1\}$. Now, since the set $\Pi_\downarrow^{(1)}$ describes a polytope, we can solve problem~\eqref{eq:general_solution} as a linear program, which gives $D_\mathrm{min} (\gamma, \Pi^{(1)}) = 0.3$. Note that the set $\Pi^{(1)}$ contains a least-informative distribution for $k=2$, i.e., $\pi^* = (0.2, 0.4, 0.2, 0.2)$, so we could have also used the result of Proposition~\ref{prop:prior_set_contains_least_informative} to solve the problem. An example of an optimal mechanism for this problem is: 
\begin{equation}
    P^* = 
    \begin{bmatrix}
    0 & \frac{1}{3} & \frac{1}{3} & \frac{1}{3} \\
    0 & 1 & 0 & 0 \\ 
    0 & 0 & 1 & 0\\
    0 & 0.3 & 0.2 & 0.5 
    \end{bmatrix}, 
\end{equation}
which achieves $D_\mathrm{min} (\gamma, \Pi^{(1)})$ with $\pi^*$ as the prior. 
\end{example}

\begin{example}
Now, suppose $\gamma = \log \, 2.5$ and we wish to find $D_\mathrm{min} (\gamma, \Pi)$ for $\Pi = \Pi^{(1)} \cup \Pi^{(2)}$, where $\Pi^{(1)}$ was defined in the previous example, and 
\begin{equation}
    \Pi^{(2)} = \{(0.3, 0.3, 0.1, 0.3), (0.29, 0.28, 0.29, 0.14), (0.05, 0.15, 0.4, 0.4)\}. 
\end{equation}
Clearly, the set $\Pi$ does not contain a least-informative distribution for $k=2$. However, each of the two sets $\Pi^{(1)}$ and $\Pi^{(2)}$ do contain a least-informative distribution ($(0.29, 0.28, 0.29, 0.14)$ is least-informative in $\Pi^{(2)}$), so it suffices to compare $D_{\mathrm{min}}(\gamma, \pi)$ for $\pi = (0.2, 0.4, 0.2, 0.2)$ and $\pi = (0.29, 0.28, 0.29, 0.14)$. By doing so, we get $D_\mathrm{min} (\gamma, \Pi) = 0.28$ which is achieved by $\pi = (0.29, 0.28, 0.29, 0.14)$. An example of an optimal mechanism for this problem is: 
\begin{equation}
    P^* = 
    \begin{bmatrix}
    1 & 0 & 0 & 0 \\
    0.1 & 0.5 & 0.4 & 0 \\ 
    0 & 0 & 1 & 0\\
    0.3 & 0.4 & 0.3 & 0 \\
    \end{bmatrix}.
\end{equation}
\end{example}

We conclude this section by stating an upper bound on $D_\mathrm{min} (\gamma, \Pi)$ that is valid for all $\Pi \neq \emptyset$. 
\begin{remark}
\label{rem:most_reliable_mechanism}
Consider a mechanism $Q^* \in \mathcal S_\gamma$ satisfying $\max_i q^*_{ij} = q^*_{jj} = \frac{e^\gamma}{n}$ for all $j \in [n]$. Then, for all $\Pi \neq \emptyset$,
\begin{align}
\begin{split}
    D_\mathrm{min} (\gamma, \Pi) &= 1 - \sup_{P \in \mathcal S_\gamma} \; \inf_{\pi \in \Pi} \sum_j p_{jj} \pi_j \\
    &\leq 1 - \inf_{\pi \in \Pi} \sum_j \frac{e^\gamma}{n} \pi_j \\
    &= 1 - \frac{e^\gamma}{n}. 
\end{split}
\end{align}
\end{remark}
As we showed in Proposition~\ref{prop:prior_set_contains_uniform}, this upper bound is attained when $\Pi$ contains the uniform distribution. The intuition behind this upper bound will be made clear in the next section. 

%% file: sections/prob3.tex
In this section, we will consider a slightly different problem. Suppose the set $\Pi$ contains all prior distributions with support of size $n$, that is, $\Pi$ is the relative interior of $\Delta^{(n-1)}$. We are given two privacy mechanisms $P, Q \in \mathcal S_\gamma$, and we want to compare the largest distortion generated by them. Thus, we want to find $D_\mathrm{max} (P)$ defined as 
\begin{equation}
    D_\mathrm{max} (P) \coloneqq \sup_{\pi \in \Pi} (1 - \sum_j p_{jj} \pi_j) = 1 - \inf_{\pi \in \Pi} \sum_j p_{jj} \pi_j,
\end{equation}
and compare it with $D_\mathrm{max} (Q)$.

\begin{theorem}
\label{theo:max_distortion}
Let $p_\mathrm{diag} = (p_{11}, \ldots, p_{nn})$ denote the vector of diagonal entries for matrix $P$. $D_\mathrm{max} (P)$ is Schur-convex and decreasing in $p_\mathrm{diag}$. Therefore, for $P,Q \in \mathcal S_\gamma$,
\begin{equation}
    q_\mathrm{diag} \prec^w p_\mathrm{diag} \quad \implies \quad D_\mathrm{max} (Q)  \leq D_\mathrm{max} (P).
\end{equation}
\end{theorem} 
\begin{IEEEproof}
We can write
\begin{align}
\label{eq:max_distortion}
\begin{split}
    D_\mathrm{max} (P) = 1 - \inf_{\pi \in \Pi} \sum_j p_{jj} \pi_j = 1 - \min_j p_{jj}. 
\end{split}
\end{align}
Note that the infimum in~\eqref{eq:max_distortion} cannot be attained since the prior attaining it is on the boundary of $\Delta^{(n-1)}$. Schur-convexity of $D_\mathrm{max} (P)$ follows from the fact that $1 - \min_j p_{jj}$ is convex and symmetric with respect to permutations of $p_\mathrm{diag}$~\cite[3.C.2]{Marshall2011}. It is also easy to see that $D_\mathrm{max}(P)$ is decreasing in every component of $p_\mathrm{diag}$, while keeping the other components constant. 
\end{IEEEproof}

Intuitively, we can view $D_\mathrm{max}(P)$ as the capacity of mechanism $P$ for generating distortion. Therefore, Theorem~\ref{theo:max_distortion} states that mechanisms with larger and more uniform diagonal entries have a lower capacity for generating distortion. Hence, if there is high uncertainty in what the true prior is, such as the case when $\Pi$ is the relative interior of $\Delta^{(n-1)}$, it is better to pick a mechanism with larger and more uniform diagonal entries to avoid large distortion.
\begin{corollary}
Let $Q^*$ be the mechanism satisfying  $\max_i q^*_{ij} = q^*_{jj} = \frac{e^\gamma}{n}$ for all $j \in [n]$. From Theorem~\ref{theo:max_distortion}, we can conclude that $D_\mathrm{max}(Q^*) \leq D_\mathrm{max}(P)$ for all $P \in \mathcal{S_\gamma}$. Furthermore, the expected distortion generated by $Q^*$ does not depend on the prior distribution. Hence, for all $\pi$, we have 
\begin{equation}
    \mathbb E_{P^*, \pi} [d(X, Y)] = 1 - \sum_j p^*_{jj} \pi_j =1 - \frac{e^\gamma}{n} \sum_j \pi_j = 1 - \frac{e^\gamma}{n}.
\end{equation}
\end{corollary}
The previous corollary and Remark~\ref{rem:most_reliable_mechanism} state that the distortion generated by mechanism $Q^*$, which has a uniform privacy cost over the symbols, has no dependency on the prior distribution. Intuitively, we can conclude that $Q^*$ is the most reliable mechanism when either there is high uncertainty in the true value of the prior (i.e., when $\Pi$ is the set of all distributions), or when the prior is not informative (i.e., when $\pi$ is the uniform distribution).   

%% file: sections/conclusion.tex
In this paper, we have studied the privacy-utility tradeoff using maximal leakage as the measure of privacy and the expected Hamming distortion as the measure of utility. In this context, we have formulated three different but related problems. First, we assumed that the prior distribution is known, and we considered the problem of finding the optimal privacy mechanism that minimizes the expected Hamming distortion subject to a maximal leakage constraint. Then, we generalized this setup to a scenario in which the prior is not exactly known, but belongs to some set of distributions. Here, we formulated a min-max problem for finding the smallest expected distortion for the worst-case prior in the set subject to a maximal leakage constraint. In our last problem, we compared privacy mechanisms in terms of the largest distortion they can create, and partially ordered them accordingly. 

Our results show that when the prior distribution is known, the optimal privacy mechanism is opportunistic in that the privacy budget is allocated only to symbols with the largest prior probabilities, while symbols with the smallest prior probabilities are completely suppressed. Roughly speaking, this is because both our utility and privacy measures can be decomposed as the sum of the corresponding per-symbol measures. More specifically, the expected Hamming distortion is a weighted sum utility: The diagonal entries in the matrix of a privacy mechanism represent the per-symbol utilities which are weighted by the prior probabilities. Similarly, maximal leakage can be thought of as the sum of the privacy costs incurred by the output symbols. Taking this interpretation into account, we may expect conceptually comparable results in studying the privacy-utility tradeoff using other privacy/utility measures that demonstrate a similar decoupling behavior. As such, the methods used in our work can be applied to studying other privacy-utility problems. 

For our second and third problems, we used majorization theory to show that priors which are more uniform, and therefore less informative, lead to larger distortion. On the other hand, privacy mechanisms that allocate the privacy budget more uniformly to the symbols generate smaller worst-case distortion. Hence, these results are valuable in that they provide general guidelines for designing high-utility privacy mechanisms. 

%% file: main.bbl
\begin{thebibliography}{10}
\providecommand{\url}[1]{#1}
\csname url@samestyle\endcsname
\providecommand{\newblock}{\relax}
\providecommand{\bibinfo}[2]{#2}
\providecommand{\BIBentrySTDinterwordspacing}{\spaceskip=0pt\relax}
\providecommand{\BIBentryALTinterwordstretchfactor}{4}
\providecommand{\BIBentryALTinterwordspacing}{\spaceskip=\fontdimen2\font plus
\BIBentryALTinterwordstretchfactor\fontdimen3\font minus
  \fontdimen4\font\relax}
\providecommand{\BIBforeignlanguage}[2]{{%
\expandafter\ifx\csname l@#1\endcsname\relax
\typeout{** WARNING: IEEEtran.bst: No hyphenation pattern has been}%
\typeout{** loaded for the language `#1'. Using the pattern for}%
\typeout{** the default language instead.}%
\else
\language=\csname l@#1\endcsname
\fi
#2}}
\providecommand{\BIBdecl}{\relax}
\BIBdecl

\bibitem{braun2009quantitative}
C.~Braun, K.~Chatzikokolakis, and C.~Palamidessi, ``Quantitative notions of
  leakage for one-try attacks,'' \emph{Electronic Notes in Theoretical Computer
  Science}, vol. 249, pp. 75--91, 2009.

\bibitem{issa2019operational}
I.~Issa, A.~B. Wagner, and S.~Kamath, ``An operational approach to information
  leakage,'' \emph{IEEE Transactions on Information Theory}, vol.~66, no.~3,
  pp. 1625--1657, 2019.

\bibitem{sara2021}
S.~Saeidian, G.~Cervia, T.~J. Oechtering, and M.~Skoglund, ``Quantifying
  membership privacy via information leakage,'' \emph{IEEE Transactions on
  Information Forensics and Security}, 2021.

\bibitem{dwork2014algorithmic}
C.~Dwork, A.~Roth \emph{et~al.}, ``The algorithmic foundations of differential
  privacy.'' \emph{Foundations and Trends in Theoretical Computer Science},
  vol.~9, no. 3-4, pp. 211--407, 2014.

\bibitem{sarwate2014rate}
A.~D. Sarwate and L.~Sankar, ``A rate-disortion perspective on local
  differential privacy,'' in \emph{2014 52nd Annual Allerton Conference on
  Communication, Control, and Computing (Allerton)}, 2014, pp. 903--908.

\bibitem{wang2016relation}
W.~Wang, L.~Ying, and J.~Zhang, ``On the relation between identifiability,
  differential privacy, and mutual-information privacy,'' \emph{IEEE
  Transactions on Information Theory}, vol.~62, no.~9, pp. 5018--5029, 2016.

\bibitem{kalantari2018robust}
K.~Kalantari, L.~Sankar, and A.~D. Sarwate, ``Robust privacy-utility tradeoffs
  under differential privacy and {Hamming} distortion,'' \emph{IEEE
  Transactions on Information Forensics and Security}, vol.~13, no.~11, pp.
  2816--2830, 2018.

\bibitem{alvim2011differential}
M.~S. Alvim, M.~E. Andr{\'e}s, K.~Chatzikokolakis, P.~Degano, and
  C.~Palamidessi, ``Differential privacy: on the trade-off between utility and
  information leakage,'' in \emph{International Workshop on Formal Aspects in
  Security and Trust}.\hskip 1em plus 0.5em minus 0.4em\relax Springer, 2011,
  pp. 39--54.

\bibitem{duchi2013local}
J.~C. Duchi, M.~I. Jordan, and M.~J. Wainwright, ``Local privacy and
  statistical minimax rates,'' in \emph{2013 IEEE 54th Annual Symposium on
  Foundations of Computer Science}, 2013, pp. 429--438.

\bibitem{kairouz2016extremal}
P.~Kairouz, S.~Oh, and P.~Viswanath, ``Extremal mechanisms for local
  differential privacy,'' \emph{The Journal of Machine Learning Research},
  vol.~17, no.~1, pp. 492--542, 2016.

\bibitem{liao2018privacy}
J.~Liao, O.~Kosut, L.~Sankar, and F.~P. Calmon, ``Privacy under hard distortion
  constraints,'' in \emph{2018 IEEE Information Theory Workshop (ITW)}, 2018,
  pp. 1--5.

\bibitem{rassouli2019optimal}
B.~Rassouli and D.~G{\"u}nd{\"u}z, ``Optimal utility-privacy trade-off with
  total variation distance as a privacy measure,'' \emph{IEEE Transactions on
  Information Forensics and Security}, vol.~15, pp. 594--603, 2019.

\bibitem{Marshall2011}
A.~W. Marshall, I.~Olkin, and B.~C. Arnold, \emph{Inequalities: Theory of
  Majorization and Its Applications}.\hskip 1em plus 0.5em minus 0.4em\relax
  Springer New York, 2011.

\end{thebibliography}
